%% file: arxiv_version.tex
\documentclass[conference]{IEEEtran}

\usepackage[utf8]{inputenc}
\usepackage[T1]{fontenc}
\usepackage{microtype}
\usepackage{graphicx}
\usepackage{tikz}
\usetikzlibrary{positioning,calc,arrows.meta}
\usepackage{booktabs}
\usepackage{tabularx}
\usepackage{listings}
\usepackage{xcolor}
\usepackage{amsmath,amssymb}
\usepackage{hyperref}
\usepackage{cleveref}
\usepackage{enumitem}
\usepackage{url}
\usepackage{comment}
\usepackage{subcaption}
\input{comments}

\hypersetup{
  colorlinks=true,
  linkcolor=blue!50!black,
  citecolor=green!40!black,
  urlcolor=blue!60!black,
}

\definecolor{codegray}{gray}{0.95}
\lstdefinestyle{dpuasm}{
  basicstyle=\ttfamily\small,
  breaklines=true,
  columns=fullflexible,
  keepspaces=true,
  showstringspaces=false,
}
\newcommand{\upmem}{\textsc{UPMEM}}
\newcommand{\llvm}{\textsc{llvm}}
\newcommand{\reg}[1]{\texttt{#1}}
\newcommand{\code}[1]{\texttt{#1}}

\title{PIMROP: A Feasibility Study of Return-Oriented Programming\\
on Commercial Processing-in-Memory Architectures}

\input{authors}

\begin{document}
\maketitle

\begin{abstract}
Processing-in-Memory places small programmable cores on the memory device, where they execute beside the DRAM banks and beyond the control-flow defenses that protect host programs. The security of this in-memory code is an attack surface that remains largely unstudied. We present PIMROP, to our knowledge the first demonstrated code-reuse attack on a commercial Processing-in-Memory ISA. We show that return-oriented programming is feasible on a real, commercially shipped PIM instruction set, the UPMEM DPU, whose DRAM Processing Units run C compiled by a dedicated LLVM toolchain. We build Galileo-for-DPU, a gadget scanner adapted to the DPU's fixed-width, 8-byte-aligned RISC encoding, and run it entirely on the UPMEM functional simulator in a reproducible Docker and VM environment. We find the runtime library that every kernel links already yields 517 non-trivial gadgets covering all five canonical ROP primitive classes, and all 13 PrIM benchmarks we scanned carry gadgets as well. We then demonstrate two verified end-to-end attacks, each reproducible on \textbf{10 of 10} runs. In the first, a stack buffer overflow in a DPU kernel overwrites the saved return address and diverts control into a routine that copies a secret into host-readable memory. In the second, we leak the same secret using a chain of reused gadgets alone, without diverting control into any such routine. We further show that two standard defenses---stack canaries and \code{\_FORTIFY\_SOURCE}-style bounds-check fortification---fail on the DPU in ways specific to the platform (misplaced canary, silently inactive wrapper), and we verify a bounds-checked DMA that stops the attack on \textbf{10 of 10} runs with no effect on benign requests. To enable further research on PIM security, we open-source Galileo-for-DPU and our testing harness at \url{https://github.com/wtegge2/ROP-in-PIM}.
\end{abstract}


\section{Introduction}
\label{sec:intro}
For decades the cost of moving data between memory and the processor has grown faster than the cost of computing on it. For today's memory-bound workloads, data movement dominates both time and energy~\cite{mutlu2022primer}. Processing-in-Memory (PIM) attacks this problem directly by placing computation in or near the memory devices~\cite{mutlu2022primer, rapid2026, tegge2026low, fang2026aespim}. The UPMEM DPU is the first commercially available, general-purpose PIM product~\cite{gomezluna2022prim}, pairing each DRAM chip with programmable in-order cores (DRAM Processing Units) that run ordinary C compiled by a dedicated LLVM toolchain. A programmable processor now lives inside the memory module and with it, a new and largely unexamined attack surface.

Proposed control-flow defenses such as $W\oplus X$/DEP, address-space layout randomization (ASLR), stack canaries, Intel CET, and ARM pointer authentication harden code running on the host CPU. None of them says anything about code executing on the DPU which runs on a different processor, with a different ISA, on the far side of the memory bus. PIM-security research so far has examined side and covert channels~\cite{bostanci2024impact}, RowHammer-style bit flips~\cite{tobah2024gadgethammer}, and confidentiality of data offloaded to untrusted memory~\cite{ghinani2025secure}, and a recent threat-landscape survey~\cite{iqbal2025redefining} explicitly identifies code-reuse against programmable NDP cores as an open concern, but the integrity of control flow on the in-memory processor itself remains undemonstrated. 

We ask the question Shacham~\cite{shacham2007geometry} first posed for the x86 and Buchanan et al.~\cite{buchanan2008good} extended to RISC, now for a real PIM ISA: \emph{does a commercially deployed Processing-in-Memory processor contain enough reusable instruction fragments to mount return-oriented programming (ROP) attacks, on a target where none of the host's control-flow defenses apply?} We focus on ROP because it is what we demonstrate end-to-end; its stackless cousin, jump-oriented programming (JOP)~\cite{bletsch2011jop}, appears here only as the indirect-jump terminators our scanner catalogs and a dispatcher we leave to future work. Because the DPU ISA is fixed-width and 8-byte aligned, we use the Buchanan RISC setting, in which gadgets are \emph{intended} instruction suffixes of real functions. This method leaves out the extra ``hidden'' gadgets Shacham obtains on the x86 by decoding the same bytes at misaligned offsets, which the DPU's alignment rules out. We develop this distinction further in \Cref{sec:bg:riscframing}. We also make two exploitation observations specific to the DPU 1) the saved return is stored as an instruction \emph{index} rather than a byte address, and 2) standard stack canaries are bypassed in two independent ways on the DPU: misplaced by the toolchain, and replayable by the attacker even when correctly placed. We show that a bounds-checked DMA helps mitigate our attack. 

We present \textbf{PIMROP}, to our knowledge the first demonstrated code-reuse (ROP) attack on a commercial Processing-in-Memory ISA. Our contributions are as follows:
\begin{enumerate}[leftmargin=*,label=\arabic*)]
\item We build \textbf{Galileo-for-DPU}, a gadget scanner for the UPMEM DPU that recovers the calling convention and indirect-branch terminators from real \code{llvm-objdump} disassembly and enumerates gadgets under the DPU's fixed-width, 8-byte-aligned RISC encoding which is the Buchanan RISC setting of intended instruction suffixes, not Shacham's x86 misaligned-decoding trick.
\item We show that Shacham's feasibility thesis~\cite{shacham2007geometry} holds on real DPU code. In the runtime library alone we discover 317 indirect-branch terminators and 517 non-trivial gadgets covering all five canonical ROP primitive classes, and we find gadgets in every one of 13 PrIM benchmarks~\cite{gomezluna2022prim}.
\item We demonstrate two verified end-to-end attacks on the functional simulator against an intentionally vulnerable DPU kernel. The first hijacks the saved return and diverts control into a routine that copies a secret into host-readable memory. The second leaks the same secret using a chain of reused library gadgets alone, without diverting control into any such routine.
\item We demonstrate the two reused control-flow primitives required for Turing-complete computation from a code-reuse chain on the DPU: a conditional branch with a runtime-chosen target and a loop with a runtime-chosen iteration count, each measured as deterministic across input variations and built from the already-linked runtime library alone.
\item We present a defense-gap analysis with a measured mitigation study. The DPU side has no ASLR and no CFI or pointer authentication, and stack canaries are ineffective in two independent ways: the toolchain places the slot on the wrong side of the frame on this upward-growing stack, and even a correctly-placed canary is bypassed 10 of 10 by a sink-free chain that preserves the DPU's fixed, host-unseeded guard value. We then verify a bounds-checked DMA that does stop the attack and identify the code change needed to make canaries effective.
\end{enumerate}

\section{Background and Motivation}
\label{sec:background}

\subsection{Processing-in-Memory and the UPMEM DPU}
\label{sec:bg:pim}
Processing-in-Memory places compute logic in or near the memory arrays so that operations run where the data already lives, avoiding the energy and latency of sending it across the memory bus~\cite{mutlu2022primer}. The UPMEM system is the first commercial, general-purpose realization~\cite{gomezluna2022prim} in which standard DDR4 DIMMs are populated with DRAM Processing Units (DPUs), simple in-order cores co-located with the DRAM banks. The programming model is host-orchestrated. A program on the host CPU allocates DPUs, copies code and input data into DPU memory, launches the DPUs, and copies results back. Each DPU runs a kernel written in C and compiled by UPMEM's LLVM-based toolchain~\cite{upmemsdk}. Within a DPU, up to 24 hardware threads (tasklets) execute concurrently. \Cref{fig:dpu-arch} shows this organization.

\subsection{Microarchitecture and Memory Hierarchy}
\label{sec:bg:arch}
\begin{figure*}[t]
\centering
\includegraphics[width=0.92\textwidth]{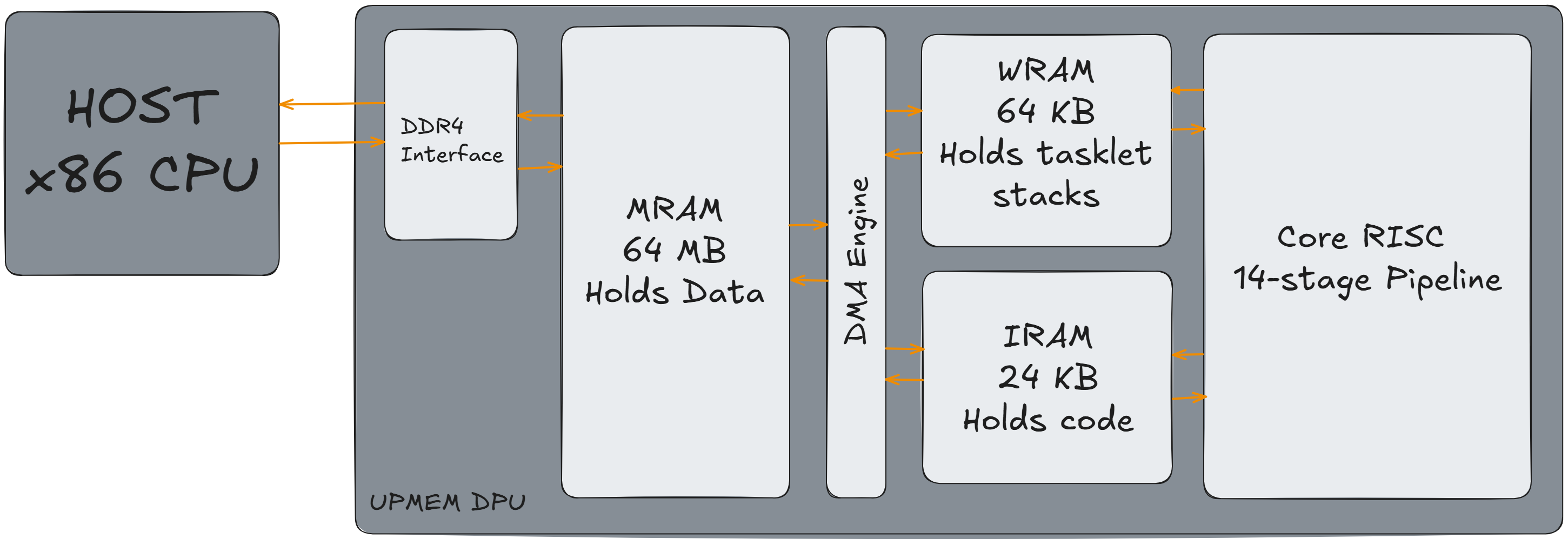}
\caption{UPMEM DPU organization. The host reaches the DPU over the DDR4 interface, and a DMA engine moves data between the MRAM bank and the on-core WRAM and IRAM that the RISC core executes from. Sizes are listed in \Cref{tab:dpu-arch}.}
\label{fig:dpu-arch}
\end{figure*}

\Cref{tab:dpu-arch} summarizes the DPU microarchitecture~\cite{gomezluna2022prim,upmemsdk}; \Cref{fig:dpu-arch} shows how the pieces connect. The calling convention (\Cref{sec:setup:callconv}) uses \reg{r23} as the link register and \reg{r22} as the frame and stack pointer. The DPU side carries none of the standard control-flow protections (\Cref{sec:eval:defense}).


\begin{table}[t]
  \centering
\caption{UPMEM DPU Architecture}
  \label{tab:dpu-arch}
  \begin{tabularx}{\linewidth}{@{}lX@{}}
    \toprule
    \textbf{Property} & \textbf{Value} \\
    \midrule
    Word size            & 32-bit \\
    Instruction width    & 64 bits (8 bytes), fixed-width, aligned \\
    IRAM (code)          & 24\,KB, up to 4096 instructions \\
    WRAM (scratchpad)    & 64\,KB --- the stack lives here \\
    MRAM (main)          & 64\,MB DRAM bank \\
    Hardware threads     & up to 24 tasklets per DPU \\
    Registers / thread   & 24 $\times$ 32-bit GPR \\
    Pipeline             & 14-stage, in-order \\
    Link register        & \reg{r23} (holds return address on \code{call}) \\
    Frame/stack pointer  & \reg{r22} \\
    CFI protections      & none (no canaries, no ASLR, no W$\oplus$X on DPU) \\
    \bottomrule
  \end{tabularx}
\end{table}

\subsection{Code-Reuse Attacks: ROP and JOP}
\label{sec:bg:rop}
Modern systems resist code-injection attacks with \emph{write-xor-execute} ($W\oplus X$, also called DEP/NX) in which memory is either writable or executable but never both, so an attacker cannot run data they inject. Code-reuse attacks sidestep this by executing code that is already present and executable in the target. Return-oriented programming (ROP)~\cite{shacham2007geometry} corrupts the stack so that a chain of short instruction sequences --- also called ``gadgets,'' each ending in a \code{ret} --- executes in sequence, with the stack supplying the next gadget's address after every return. Prior work ~\cite{checkoway2010returnless} showed that any indirect branch suffices, and the terminator need not be a \code{ret}. Jump-oriented programming (JOP)~\cite{bletsch2011jop} drops the stack entirely and creates chains through indirect jumps and a central dispatcher gadget. 
The central claim underlying both of these is that a sufficiently large body of code contains enough reusable gadgets to assemble arbitrary, even Turing-complete, computation~\cite{shacham2007geometry}. Shacham identifies five canonical primitive classes a gadget set must span to be computationally universal: \emph{load-constant}, \emph{load-from-memory}, \emph{store-to-memory}, \emph{arithmetic/logic}, and \emph{control-flow} (data-dependent branches and loops)~\cite{shacham2007geometry,buchanan2008good}. Showing that a target realizes all five is therefore the standard feasibility criterion, and the property we measure for the DPU in \Cref{sec:eval:primitives}.

\subsection{Why Fixed-Width RISC (Galileo vs.\ Buchanan)}
\label{sec:bg:riscframing}
Shacham's Galileo algorithm~\cite{shacham2007geometry} exploits a peculiarity of the x86. Decoding the same bytes at a different offset yields different, unintended instructions because instructions are variable-length and unaligned, so a binary hides far more gadgets than its intended instruction stream contains. The DPU ISA removes this entirely. Instructions are fixed-width and 8-byte aligned, so there are no ``hidden'' instructions to uncover. This means every gadget must begin at a real instruction boundary. This places our work in the setting of Buchanan et al.~\cite{buchanan2008good}, who generalized ROP to fixed-width aligned RISC (SPARC), where gadgets are intended instruction suffixes of real functions.

\subsection{Stack Layout and Overflow Mechanics}
\label{sec:design:layout}
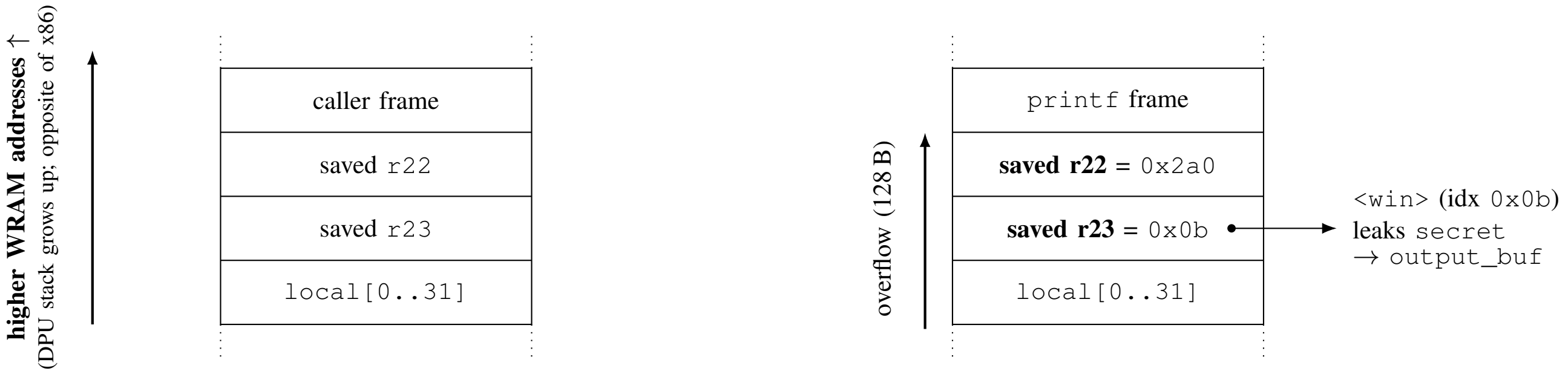
\begin{figure*}[t]
\centering
\begin{tikzpicture}[font=\footnotesize, >={Latex[length=1.6mm]}]
  \draw[->, thick] (-1.4,0) -- (-1.4,3.0);
  \node[rotate=90, anchor=south, align=center] at (-1.65,1.5)
    {\textbf{higher WRAM addresses} $\uparrow$\\
     \scriptsize(DPU stack grows up; opposite of x86)};

  \begin{scope}
    \draw (0,0) rectangle (3.4,2.8);
    \draw (0,0.7) -- (3.4,0.7);
    \draw (0,1.4) -- (3.4,1.4);
    \draw (0,2.1) -- (3.4,2.1);
    \draw[dotted] (0,2.8) -- (0,3.2);  \draw[dotted] (3.4,2.8) -- (3.4,3.2);
    \draw[dotted] (0,0)   -- (0,-0.4); \draw[dotted] (3.4,0)   -- (3.4,-0.4);
    \node at (1.7,0.35) {\texttt{local[0..31]}};
    \node at (1.7,1.05) {saved \texttt{r23}};
    \node at (1.7,1.75) {saved \texttt{r22}};
    \node at (1.7,2.45) {caller frame};
    \node[anchor=north] at (1.7,-0.5) {\textbf{(a)} baseline non-leaf frame};
  \end{scope}

  \begin{scope}[xshift=8cm]
    \draw (0,0) rectangle (3.4,2.8);
    \draw (0,0.7) -- (3.4,0.7);
    \draw (0,1.4) -- (3.4,1.4);
    \draw (0,2.1) -- (3.4,2.1);
    \draw[dotted] (0,2.8) -- (0,3.2);  \draw[dotted] (3.4,2.8) -- (3.4,3.2);
    \draw[dotted] (0,0)   -- (0,-0.4); \draw[dotted] (3.4,0)   -- (3.4,-0.4);
    \node at (1.7,0.35) {\texttt{local[0..31]}};
    \node at (1.7,1.05) {\textbf{saved r23} = \texttt{0x0b}};
    \node at (1.7,1.75) {\textbf{saved r22} = \texttt{0x2a0}};
    \node at (1.7,2.45) {\texttt{printf} frame};
    \node[anchor=north] at (1.7,-0.5) {\textbf{(b)} during overflow in \code{parse\_request}};
    \draw[->, thick] (-0.3,-0.05) -- (-0.3,2.1);
    \node[rotate=90, anchor=south] at (-0.5,1.05) {overflow (128\,B)};
    \fill (3.05,1.05) circle (1.3pt);
    \draw[->] (3.05,1.05) -- (4.2,1.05);
    \node[anchor=west, align=left] at (4.25,1.05)
      {\texttt{<win>} (idx \texttt{0x0b})\\[1pt]
       leaks \texttt{secret}\\[-1pt]$\to$ \texttt{output\_buf}};
  \end{scope}
\end{tikzpicture}
\caption{DPU stack layout and overflow mechanic (addresses grow upward, opposite of x86). \textbf{(a)} baseline non-leaf frame. \textbf{(b)} overflow moment in \code{parse\_request} (\Cref{sec:eval:e2e}): the unbounded \code{mram\_read} runs upward from the \code{local} buffer and overwrites the saved \texttt{\{r22,r23\}} pair, so the epilogue transfers to the attacker-written value.}
\label{fig:stack-layout}
\end{figure*}

In our running-example vulnerable function (\code{parse\_request}, detailed in \Cref{sec:design:vuln}), the saved \code{\{r22,r23\}} pair sits just past the vulnerable buffer, so a linear overflow reaches it once the buffer is full. WRAM has no address randomization, so the buffer, stack, and every gadget and sink address are fixed and known offline from the binary. Only the in-frame offset to the saved pair varies between builds. The saved pair is stored with \reg{r23} in the low word and \reg{r22} in the high word---the opposite of the naive guess---so the overflow must place the jump target in the low word and a valid frame pointer in the high word. Rather than aim at one slot, we saturate the region past the buffer with copies of the pair so that whichever 8-byte-aligned slot ends up holding the saved return receives a valid \{\reg{r23},\reg{r22}\} (\Cref{sec:design:payload}). \Cref{fig:stack-layout} shows the baseline frame and the overflow moment side by side.



\section{Threat Model}
\label{sec:threat}
We consider a deployed \upmem{}-style PIM system in which a host offloads work to DPU kernels. A concrete setting is multi-tenant DPU inference: tenant A's request shape parameterizes an \code{mram\_read} on a DPU shared with tenant B. A single missing length clamp hands A a chain assembled from the shared runtime library (\Cref{sec:eval:e2e}) that leaks B's secret into MRAM A can read back, with no DPU-code injection and no knowledge of B's kernel source. Our trust boundary is the DPU itself: we treat the host CPU, OS, and orchestration runtime as trusted, and ask only what an adversary can achieve on the in-memory processor. Prior PIM-security work considers adjacent threats (timing/covert channels~\cite{bostanci2024impact}, RowHammer~\cite{tobah2024gadgethammer}, confidentiality of data in untrusted memory~\cite{ghinani2025secure,arafin2020pimsecurity}) but not control-flow integrity on the DPU, which is our focus.

\paragraph{Attacker capabilities} The attacker (i) supplies input to a DPU kernel that contains a memory-safety bug --- our running example is an unbounded \code{mram\_read} that copies an attacker-controlled number of bytes into a fixed-size WRAM buffer --- and (ii) knows the DPU runtime library the target kernel links against, which is a reasonable assumption. The SDK and its runtime sources are public~\cite{upmemsdk}, so an attacker can rebuild \texttt{librt.a} byte-identically to a victim deployment and enumerate every runtime gadget offline. We do not assume the attacker knows the proprietary user kernel. In fact, the sink-free leak of \Cref{sec:eval:sinkfree} performs the exfiltrating write using only reused library gadgets, so the kernel it runs against need only link the DPU runtime, not share any user code with the attacker. Because the DPU provides no address-space layout randomization (ASLR), gadget and sink addresses are static and computed offline from the known runtime binary (IRAM is fixed at \code{0x80000000} and further explained in \Cref{sec:bg:arch}).

\paragraph{Restrictions} The attacker cannot inject or modify DPU code. All executed instructions already exist in the kernel or in the runtime library it links. This is a code-reuse setting, consistent with a $W\oplus X$-style assumption on instruction memory, and is what distinguishes the attack from classical shellcode injection.

\paragraph{Goals} The adversary's goal is to chain reused gadgets into arbitrary, Turing-complete computation. We establish that the necessary gadgets and primitives exist, demonstrate multi-gadget chaining of control flow, and further demonstrate a fully general, sink-free chain --- one that performs the exfiltration from reused gadgets alone, rather than by routing control into an existing function (\Cref{sec:eval:sinkfree}).

\paragraph{Out of scope} Host-side compromise, physical and fault attacks, and data-oriented corruption such as RowHammer~\cite{tobah2024gadgethammer} are orthogonal and out of scope. We assume both the host and the channel by which code and data are loaded onto the DPU are trusted.

\paragraph{Bug reachability in deployed PIM software} The attack of \Cref{sec:eval:e2e} needs only one ingredient: a destination-side overflow of a WRAM buffer through an unbounded \code{mram\_read}. The reason it is dangerous is not that it is common today but that nothing on the DPU catches it once it appears (\Cref{sec:eval:defense}). An audit of all 48 \code{mram\_read} call sites in the 16 PrIM benchmarks~\cite{gomezluna2022prim}: 42 pass a compile-time constant length, and the remaining 6 clamp their runtime-computed length at the source. Zero of 48 sites admit the destination-side overflow we exploit---the bug class is \emph{latent} in current PrIM code rather than endemic. The clamp-at-every-call discipline PrIM maintains is a per-site programmer convention with no backstop, and as PIM moves toward multi-tenant and ML-inference deployments~\cite{yuksel2026memcentric,ghinani2025secure}, that discipline becomes harder to guarantee. The exploit characterizes a durable platform property rather than a specific bug.

\section{Related Work}
\label{sec:related}

\subsection{Code-Reuse Attacks: Lineage and Setting}
\label{sec:related:lineage}
Shacham~\cite{shacham2007geometry} introduced ROP on the x86; subsequent work generalized the technique to fixed-width aligned RISC~\cite{buchanan2008good}, removed the dependence on \code{ret}~\cite{checkoway2010returnless}, introduced jump-oriented variants~\cite{bletsch2011jop}, demonstrated the method on RISC-V~\cite{jaloyan2020rvrop,gu2020norisc}, and automated gadget discovery and chaining through symbolic execution for RISC-V and ARM64~\cite{cloosters2022riscyrop}. Production scanners such as ROPgadget~\cite{salwan_ropgadget} already handle aligned ISAs one instruction at a time. Beyond the CPU, Guo et al.~\cite{guo2024gpu} and Roels et al.~\cite{roels2025cuda} show that GPUs and CUDA SASS admit code-reuse execution with no deployed defenses, and defensive work has followed on the GPU side as well~\cite{santos2026warpguard,santos2026gerrymandering}. Our method---disassembling and scanning backward from indirect-branch terminators for intended instruction suffixes---is the established approach for a fixed-width aligned RISC target and is not a technical contribution. What is new is the domain: to our knowledge no prior work poses Shacham's feasibility question on a Processing-in-Memory ISA or demonstrates a control-flow hijack on the memory side of the bus, where the host's control-flow defenses do not reach. The GPU accelerator-side results are the closest methodological precedents, but they target host-launched compute accelerators, whereas on PIM the host CPU is the attack's victim rather than its launcher.

\subsection{PIM Security}
\label{sec:related:pim}
Prior PIM-security work targets different mechanisms than code reuse. Bostanci et al.~\cite{bostanci2024impact} build timing channels amplified by PIM operations, and Tobah et al.~\cite{tobah2024gadgethammer} flip nested pointers with RowHammer, a data-corruption primitive that shares the word gadget but not the mechanism of reused instruction sequences. Ghoflsaz Ghinani et al.~\cite{ghinani2025secure} propose a defense for computing on untrusted in-memory hardware, and surveys by Arafin and Lu~\cite{arafin2020pimsecurity}, Yuksel et al.~\cite{yuksel2026memcentric}, and Waingankar et al.~\cite{iqbal2025redefining} catalog the PIM threat landscape; the last of these explicitly names code-reuse against programmable NDP cores as an open concern. None of this work demonstrates a control-flow-hijacking or code-reuse attack on a PIM ISA, which is the gap this paper closes with a quantified gadget inventory and a working exploit.

\section{Methodology and Experimental Setup}
\label{sec:setup}

\subsection{Reproducible Environment}
\label{sec:setup:env}
We run on the \upmem{} DPU functional simulator (no physical PIM hardware) using the \upmem{} SDK 2023.1.0 from the CMU-SAFARI mirror~\cite{upmemsdk}, driven by the stock \code{dpu-upmem-dpurte-clang} and \code{llvm-objdump~12.0.0}. Our aarch64 host runs x86-64 SDK binaries under QEMU user-mode (scan pipeline in a container, end-to-end exploit in a full-system x86-64 guest), which affects tool speed only, not the decoded instruction stream.

\subsection{Target Binaries}
\label{sec:setup:targets}
We study two targets: our vulnerable kernel \code{vuln\_dpu} (\Cref{sec:design:vuln}) and the DPU runtime library \texttt{librt.a} (6{,}711 instructions; profiling build \texttt{librt\_p.a} at 8{,}264) that the SDK links into every DPU kernel~\cite{upmemsdk}. Because \code{ld.lld} is unstable under QEMU, we scan the pre-built archives directly rather than freshly linking. The disassembly is byte-identical to a native run. LTO archives ship as \llvm{} bitcode with no native code~\cite{lattner2004llvm,llvmlto}. PrIM ships 16 benchmarks. Our source-level \code{mram\_read} audit (\Cref{sec:threat}) covers all 16, while the gadget scan (\Cref{sec:eval:prim}) and fortification diff (\Cref{sec:eval:overhead}) cover the 13 and 12 benchmarks respectively that build cleanly under our simulator toolchain.



\subsection{Recovering the Calling Convention}
\label{sec:setup:callconv}
The precondition for any return hijack is that a function's return address lives on attacker-reachable, writable memory~\cite{alephone1996smashing}. We recovered the DPU calling convention directly from disassembly (shown in \Cref{lst:callconv}), and it satisfies this condition. On the DPU the frame pointer is \reg{r22} and the link register is \reg{r23}, which holds the return address on \code{call}. On entry, a non-leaf function spills the caller's \code{\{r22,r23\}} register pair to its WRAM stack frame (\code{sd r22, 16, d22}) and then advances \reg{r22} to grow the frame. Callee-saved pairs such as \code{\{r14,r15\}} are saved alongside it. On exit, the epilogue reverses this: it reloads the saved pair from the stack (\code{ld d22, r22, -8}) and returns by branching to the restored link register (\code{jump r23}). An attacker who can write past a stack buffer controls the branch target because WRAM is writable and the saved \reg{r23} is read back off the stack immediately before that indirect \code{jump r23}. This shows the classic ROP precondition on the memory side of the bus and is consistent with the upmem calling convention~\cite{upmemsdk}.

Two details of this convention shape the payload (\Cref{sec:eval:e2e}). First, the pair is stored as \{\reg{r23} in the low word, \reg{r22} in the high word\}, so the overflow must place the jump target in the low word and a valid frame pointer in the high word. Second, \reg{r23} is not a byte address but an instruction index, so writing a raw code address lands control at the wrong place. Finally, only the callee-saved pairs \reg{r14}--\reg{r23} are spilled to the stack and thus recoverable by an attacker, which narrows the set of library routines a chain can thread through but does not block it (\Cref{sec:eval:e2e}).

\begin{lstlisting}[float=t,captionpos=b,columns=fixed,breaklines=false,%
basicstyle=\ttfamily\footnotesize,caption={Observed DPU prologue/epilogue: the
saved link register r23 is spilled to the WRAM stack and reloaded before the
indirect `jump r23' that implements return.},label={lst:callconv}]
; prologue -- save {r22,r23}, grow frame
sd   r22, 16, d22   ; spill {r22,r23} to stack
add  r22, r22, 24   ; grow frame
sd   r22, -24, d14  ; save {r14,r15}

; epilogue -- restore and return
ld   d14, r22, -24  ; restore {r14,r15}
ld   d22, r22, -8   ; reload {r22,r23} (ret addr)
jump r23            ; return == ROP terminator
\end{lstlisting}

\subsection{The Galileo-for-DPU Gadget Scanner}
\label{sec:setup:scanner}
\Cref{fig:galileo} gives our scanner. It follows Shacham's Galileo~\cite{shacham2007geometry} in scanning backward from every terminator for straight-line suffixes but adapts to the DPU's fixed-width aligned ISA in four ways: (i) it steps one aligned 8-byte instruction at a time rather than one byte at a time (no misaligned decodes, no trie); (ii) it keys on register-indirect \code{jump r23} (ROP terminator) and \code{jump rN} with $N\ne 23$ (JOP terminator) rather than the \code{ret} byte \code{c3}, ignoring direct \code{jump 0xADDR}; (iii) it parses \code{llvm-objdump} text, where instructions already sit at true boundaries, rather than decoding raw bytes; (iv) it grows gadgets backward only while the window stays 8-byte contiguous and contains no control-flow diverter before the terminator (including the DPU's \emph{embedded conditional branches}---ALU ops that also carry a \code{<cond>,<target>} operand). Gadgets are classified by capability and flagged chainable when they either reload \reg{r23} or are JOP. The backward walk is bounded by \textsc{maxlen}, which \Cref{sec:eval:inventory} sweeps.


\begin{figure}[t]
\centering
\begin{lstlisting}[basicstyle=\footnotesize\ttfamily,columns=fullflexible,%
breaklines=false,aboveskip=2pt,belowskip=2pt]
Algorithm Galileo-for-DPU:
  decode "llvm-objdump -d" text into an aligned
    instruction list I[0..n-1]   (8-byte stride)
  for i in 0 .. n-1 do:
    if I[i] is "jump rN" (register-indirect) then:
      BuildBack(i)

Procedure BuildBack(terminator index t):
  for len in 1 .. MAXLEN do:
    s := t - (len - 1)
    if s < 0:                         return
    if I[s..t] not 8-byte contiguous: return
    if any I[s..t-1] diverts flow:    return
    emit gadget I[s..t]; classify its body;
      mark chainable if it reloads r23 or is JOP
\end{lstlisting}
\caption{The Galileo-for-DPU gadget-scanning algorithm; scanner details and its four departures from Shacham's byte-recursive Galileo are in \Cref{sec:setup:scanner}.}
\label{fig:galileo}
\end{figure}

\section{Attack Design and Implementation}
\label{sec:design}

\subsection{The Vulnerable DPU Kernel}
\label{sec:design:vuln}
Our vulnerable kernel \code{vuln\_dpu} isolates the single bug this paper exploits: \code{parse\_request} uses \code{mram\_read} to DMA an attacker-controlled number of bytes into a fixed 32-byte WRAM buffer with no bounds check. The routine is marked \code{noinline} (to force its own spilled frame rather than inlining into \code{main}, which the \upmem{} runtime handles specially~\cite{upmemsdk}) and is non-leaf through its call to \code{printf}, so the calling convention of \Cref{sec:setup:callconv} spills \{\reg{r22},\reg{r23}\} just past the buffer. A long request overwrites the saved return; the epilogue's \code{jump r23} then transfers to the attacker-written address.

Two host-readable witnesses record the outcome: \code{returned\_cleanly} (a \code{\_\_host}-attributed byte~\cite{upmemsdk} set to \code{1} only on a clean epilogue) and \code{output\_buf} (MRAM scratch). The exfiltration target \code{win()}---a \code{ret2libc}-style sink that reads a host-planted \code{secret} (\code{0x00C0FFEE}) and writes it with a \code{"LEAK"} marker into \code{output\_buf}---is unreachable on any legitimate path, so landing control there demonstrates an attacker-chosen effect rather than a crash.

\subsection{Payload Construction}
\label{sec:design:payload}
Two values in each saved-pair copy must be encoded carefully. The low word is the branch target, but \reg{r23} on the DPU holds an instruction index: the byte offset from the \code{0x80000000} IRAM base divided by eight, not a byte address. Writing a raw byte address sends control to eight times the intended location and faults. The high word is \reg{r22}, the restored frame pointer, which must be a valid low-stack WRAM address (such as \code{0x2a0}) so the sink runs with headroom. With no address randomization both are fixed and are computed offline from the disassembly.

\subsection{Chaining via the Calling Convention}
\label{sec:design:chaining}
Multi-gadget chaining needs no classical stack-pivot gadget. The epilogue every non-leaf function uses (\Cref{lst:callconv}) reloads \{\reg{r22},\reg{r23}\} from the WRAM stack before branching, so the attacker-controlled payload sets \reg{r22} on every return and the next hop's frame pointer is already in place, making the calling convention itself the pivot.


\section{Evaluation}
\label{sec:eval}

\subsection{Gadget Inventory}
\label{sec:eval:inventory}
We ran Galileo-for-DPU, described in \Cref{sec:setup:scanner}, on two real targets: the DPU runtime library \texttt{librt.a} (6{,}711 instructions across 260 functions, 315 \code{jump r23} + 2 JOP terminators, 517 non-trivial gadgets, 221 chainable) and the profiling build \texttt{librt\_p.a} (8{,}264 instructions, 247 + 2 terminators, 222 non-trivial, 808 chainable). The LTO archives \texttt{librtlto*.a} are \llvm{} bitcode, not native code, and are excluded.

In \texttt{librt.a}, roughly 47 register-indirect terminators per 1000 instructions falls in the same range the RISC-ROP precedent reports for SPARC and ARM libraries~\cite{buchanan2008good,checkoway2010returnless}, so this single library already exceeds Shacham's feasibility threshold before any application code is linked. The two JOP terminators (\code{jump r1}, \code{jump r3}) are enough to exhibit a JOP sink but too thin for a full dispatcher. Gadgets are short, as code reuse requires: most are only one to three instructions long, and none exceeds five. Nearly all are return-type, with a small JOP minority. Their capability mix covers what a flexible chain needs: arithmetic, memory load and store, register move, and constant loading. These counts are stable across optimization levels.

\subsection{Primitive Coverage}
\label{sec:eval:primitives}
We next ask which of Shacham's canonical primitive classes are realizable from the gadgets found in \texttt{librt.a}, giving one representative gadget per class in \Cref{fig:primitives}.

\begin{figure*}[t]
\centering
\begin{subfigure}[t]{0.195\textwidth}
  \centering
  \begin{tabular}{@{}l@{}}
  \toprule
  \code{move r0,0}       \\
  \code{sw r0,0,0}       \\
  \code{ld d22,r22,-8}   \\
  \code{jump r23}        \\
  \bottomrule
  \end{tabular}
  \caption{Load constant\\ (71 gadgets)}
  \label{fig:prim:loadconst}
\end{subfigure}\hfill
\begin{subfigure}[t]{0.195\textwidth}
  \centering
  \begin{tabular}{@{}l@{}}
  \toprule
  \code{ld d22,r22,-8}   \\
  \code{jump r23}        \\
  \multicolumn{1}{@{}c@{}}{\scriptsize(in \code{\_\_memcpy\_mw})} \\
  \\
  \bottomrule
  \end{tabular}
  \caption{Load memory\\ (117 gadgets)}
  \label{fig:prim:loadmem}
\end{subfigure}\hfill
\begin{subfigure}[t]{0.195\textwidth}
  \centering
  \begin{tabular}{@{}l@{}}
  \toprule
  \code{sw r0,0,0}       \\
  \code{ld d22,r22,-8}   \\
  \code{jump r23}        \\
  \\
  \bottomrule
  \end{tabular}
  \caption{Store memory\\ (82 gadgets)}
  \label{fig:prim:storemem}
\end{subfigure}\hfill
\begin{subfigure}[t]{0.195\textwidth}
  \centering
  \begin{tabular}{@{}l@{}}
  \toprule
  \code{lsl r1,r1,2}     \\
  \code{lw r1,r1,0}      \\
  \code{jump r1}         \\
  \multicolumn{1}{@{}c@{}}{\scriptsize(in \code{printf})} \\
  \bottomrule
  \end{tabular}
  \caption{Arith / logic\\ (286 gadgets)}
  \label{fig:prim:arith}
\end{subfigure}\hfill
\begin{subfigure}[t]{0.195\textwidth}
  \centering
  \begin{tabular}{@{}l@{}}
  \toprule
  \code{sw r15,0,r0}     \\
  \code{move r0,r14}     \\
  \code{ld d22,r22,-8}   \\
  \code{jump r23}        \\
  \bottomrule
  \end{tabular}
  \caption{Control flow\\ (3 branch + 1 \code{ret2loop})}
  \label{fig:prim:cf}
\end{subfigure}
\caption{Shacham primitive coverage in \texttt{librt.a}. All five canonical classes are realized by the runtime library alone; counts are the total non-trivial gadgets found per class, and each panel shows one representative example gadget (gadgets within a class vary in instructions and operands).}
\label{fig:primitives}
\end{figure*}

All five canonical classes are realized. Of the 221 chainable gadgets, 217 also carry a data capability---217 with load-mem, 9 with store-mem, 13 with arith, and 4 with load-const---so chainable terminators and data operations coexist in the same gadgets rather than requiring separate chaining and data gadgets. The scan also confirms the attack-design claim of \Cref{sec:design:chaining}. No gadget writes \reg{r22} before a terminator at any scan-window size we tried. Every \reg{r22} write in \texttt{librt.a} is a function-entry frame advance (\code{add r22, r22, imm}), never epilogue-adjacent. Chaining is still feasible because the standard DPU epilogue reloads \{\reg{r22},\reg{r23}\} from the stack, so the attacker's payload threads \reg{r22} on every return without any pivot gadget.

\subsection{End-to-End Control-Flow Hijack and Secret Leak}
\label{sec:eval:e2e}
We verified the full attack end-to-end on the functional simulator against \code{vuln\_dpu} (\Cref{sec:design:vuln}). A crafted 128-byte request overruns the 32-byte buffer and redirects \code{parse\_request}'s epilogue into \code{win()}, which exfiltrates the host-planted \code{secret} (\code{0x00C0FFEE}) into \code{output\_buf}; \Cref{fig:stack-layout}(b) shows the WRAM stack at the moment of the overflow.

\paragraph{A DPU-specific exploitation detail} The saved return register is not a byte address. The DPU stores and consumes it as an instruction index which means it is the byte offset from the \code{0x80000000} IRAM base divided by eight. The hijack payload must therefore write \code{win}'s instruction index into the saved-return slot rather than its byte address, since a raw byte address lands at eight times the intended target and faults. This is a concrete, reusable finding for exploiting this ISA. It does not weaken the attack because addresses are static with no ASLR so the index is known offline, but it is required to land control precisely.

\paragraph{Determinism} With no ASLR the exploit is fully repeatable. The single static payload mentioned above leaked the secret on \textbf{10 of 10} runs.

\paragraph{A multi-gadget chain} A true multi-hop chain through the \reg{r22}-threaded reload also succeeds. The overflow sets the saved return to a reused gadget whose tail is the standard epilogue (\Cref{lst:callconv}). That gadget reads the next \{\reg{r22},\reg{r23}\} pair from an attacker-planted WRAM node and jumps on, reaching either \code{\_\_sys\_end} for a clean halt or the exfiltration sink. This establishes gadget \emph{chaining}, not merely a single hijack.

\subsection{A sink-free exfiltration} 
\label{sec:eval:sinkfree}
A sink-free chain performs the entire exfiltration from reused gadgets alone, without routing control into any pre-existing function such as \code{win()}. Two reused gadget tails suffice, threaded through the \reg{r22}-reload mechanism. The first is a \emph{setter}: a tail of a \code{printf} runtime helper that loads the callee-saved pair \{\reg{r18},\reg{r19}\} from the attacker-controlled frame and sets them to \code{\&secret} and \code{output\_buf}. The second is a \emph{DMA} tail from the runtime's \code{\_\_memcpy\_mw}, whose \code{sdma} copies the secret word from WRAM to host-readable MRAM. After the copy, the chain threads to \code{\_\_sys\_end} for a clean halt. On the simulator, the secret reaches \code{output\_buf} on \textbf{10/10} runs and the kernel halts cleanly on \textbf{10/10}. Rebuilding with \code{win()} absent gives the same \textbf{10/10}, confirming that reused gadgets alone perform the write. This chain works because the kernel links the runtime copy helpers, which is true of any kernel using \code{memcpy} or \code{memmove}. The result realizes the tenant-A/tenant-B scenario of \Cref{sec:threat}: a payload crafted against the shared runtime leaks a secret from the victim's WRAM without knowing the victim's source or injecting DPU code.

\subsection{Data-Dependent Control Flow}
\label{sec:eval:control-flow}
The chains in \Cref{sec:eval:e2e} execute a fixed sequence of gadgets, so their behavior is independent of any value computed at run time. Arbitrary rather than fixed-chain computation requires two more primitives on top of the data primitives of \Cref{sec:eval:primitives}: a conditional branch whose target is runtime-chosen and a loop whose iteration count is runtime-chosen. We realize both from reused library code.


\paragraph{Data-dependent conditional branch (store-into-chain)} The DPU threads its chain through the epilogue reload (\Cref{lst:callconv}). A reused store gadget that writes a runtime-computed value into that reload slot before the epilogue reads it yields a jump target chosen at run time, with no pivot gadget required. We realize this with a reused tail of \code{strsep}, feeding it the byte sum \code{parse\_request} already computes. We verified the mechanism two ways, each with three input values and reaching the expected outcome on 3/3 runs per value. First, we routed the runtime value through the DMA tail to \code{output\_buf} so the host could read it back directly, confirming the store wrote the correct runtime value. Second, we placed the same runtime value in the reload slot so the epilogue dispatched control to it as an instruction index, confirming the runtime value actually steers control flow. The branch target is never a constant in the payload.

\paragraph{Data-dependent loop (ret2loop into \code{strcat})} Rather than synthesize a loop from a counter and a predicate gadget, we re-enter an existing library loop mid-function. \code{strcat}'s inner copy loop iterates until the source byte is null, then returns through the standard epilogue back into our chain. A setter gadget loads the source address with a runtime value, so varying that address varies the iteration count. A DMA gadget then copies the destination region to \code{output\_buf} for the host to observe. Across input strings of varying lengths, the measured iteration count matched \code{strlen(src)} on 3/3 runs per input. The loop bound is never a constant in the payload.

These two reused primitives, together with the data primitives of \Cref{sec:eval:primitives}, supply the computational building blocks of a Turing-complete gadget set.

\subsection{Gadget Density on Real Workloads}
\label{sec:eval:prim}
To show that the attack surface is not an artifact of our demo kernel, we compiled and scanned 13 benchmarks from the PrIM suite~\cite{gomezluna2022prim}, the standard UPMEM DPU workload set reported in the PrIM suite. Every one contains register-indirect terminators and non-trivial gadgets, ranging from 8 to 49 terminators per 1000 instructions. These counts cover the application kernels on their own, and because every DPU program links the runtime the effective corpus for any of these workloads is its own gadgets together with the 517 non-trivial gadgets in \texttt{librt} reported in \Cref{sec:eval:inventory}. Gadget availability therefore holds across real PIM software and not just our target.

\section{Discussion}
\label{sec:discussion}
\subsection{Interpreting the Results}
\label{sec:disc:interp}
Shacham's feasibility thesis carries over to the DPU: the runtime library alone realizes all five canonical primitive classes and furnishes enough control for a working hijack that leaks a secret, with no code injection and nothing beyond the standard runtime required. Combined with the two reused control-flow primitives of \Cref{sec:eval:control-flow}---a data-dependent branch and a data-dependent loop---these are the building blocks of a Turing-complete gadget set; what we do not demonstrate is an automated synthesizer that composes them into arbitrary programs (\Cref{sec:future}). Because the DPU's fixed-width aligned ISA has no misaligned decodes, our scanner works in a strictly smaller search space than Shacham's x86 setting, so the gadget supply we report is a lower bound.

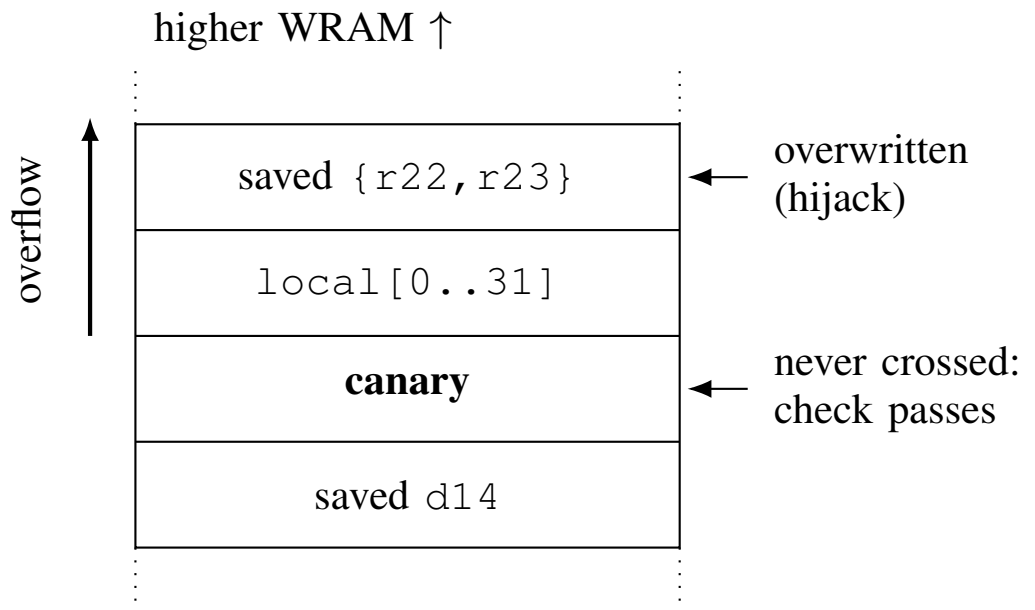
\begin{figure}[t]
\centering
\begin{tikzpicture}[font=\footnotesize, >={Latex[length=1.6mm]}]
  \draw (0,0) rectangle (3.6,2.8);
  \draw (0,0.7) -- (3.6,0.7);
  \draw (0,1.4) -- (3.6,1.4);
  \draw (0,2.1) -- (3.6,2.1);
  \draw[dotted] (0,2.8) -- (0,3.2);  \draw[dotted] (3.6,2.8) -- (3.6,3.2);
  \draw[dotted] (0,0)   -- (0,-0.4); \draw[dotted] (3.6,0)   -- (3.6,-0.4);
  \node at (1.8,0.35) {saved \texttt{d14}};
  \node at (1.8,1.05) {\textbf{canary}};
  \node at (1.8,1.75) {\texttt{local[0..31]}};
  \node at (1.8,2.45) {saved \texttt{\{r22,r23\}}};
  \node[anchor=west] at (0,3.42) {higher WRAM $\uparrow$};
  \draw[->,thick] (-0.3,1.4) -- (-0.3,2.85);
  \node[rotate=90,anchor=south] at (-0.5,2.1) {overflow};
  \draw[->] (4.05,1.05) -- (3.65,1.05);
  \node[anchor=west,align=left] at (4.1,1.05) {never crossed:\\ check passes};
  \draw[->] (4.05,2.45) -- (3.65,2.45);
  \node[anchor=west,align=left] at (4.1,2.45) {overwritten\\ (hijack)};
\end{tikzpicture}
\caption{Why stack canaries fail on the DPU: the upward-growing frame leaves the canary on the wrong side of the overflow (\Cref{sec:eval:defense}).}
\label{fig:canary}
\end{figure}

\subsection{Defense-Gap Analysis}
\label{sec:eval:defense}

We probed each standard mitigation on the DPU side. ASLR is absent because code lives at a fixed IRAM base of \code{0x80000000} and the WRAM stack is deterministic, so the exploit reproduced 10 of 10 runs from a single hard-coded payload and every gadget and sink address is known offline. The DPU also lacks any CFI or pointer-authentication. The net result is that none of the host-side protections in \Cref{tab:dpu-arch} has a working counterpart on the DPU, which is exactly why the overflow discussed in \Cref{sec:eval:e2e} is exploitable with a single static payload.

Stack canaries fail in two separate ways. First they are unavailable out of the box: the compiler front end accepts \code{-fstack-protector-all} and emits canary checks, but linking fails on undefined \code{\_\_stack\_chk\_guard} and \code{\_\_stack\_chk\_fail}, so a stack-protected kernel cannot be built at all. We removed this obstacle with a minimal runtime that defines a guard word and a \code{\_\_stack\_chk\_fail} handler that halts. A protected kernel then links. Even so, the canary does not stop the attack: refiring the same exploit against the hardened kernel still leaks the secret 10 of 10 while the canary trips 0 of 10 (\Cref{tab:mitigation}). The reason is architectural. The DPU stack frame grows toward higher addresses, so a callee saves its return pair \code{\{r22,r23\}} above its local buffer whereas the compiler places the canary below it. A linear overflow DMAs upward from the buffer into the saved return and never touches the canary (\Cref{fig:canary}). This placement is not an artifact of one build: we swept 144 configurations of the vulnerable kernel (six optimization levels, six buffer sizes from 32 to 1024 bytes, with and without frame-pointer omission, at 1 or 4 tasklets), and every build placed the canary slot at the lowest offset of the frame. We therefore treat the misplacement as a property of the toolchain port, requiring a backend fix rather than per-kernel source changes. Specifically, it belongs in an \llvm{} backend pass over the DPU stack-protector slot ordering; implementing and upstreaming the pass is left as future work.

We also verified one mitigation that does hold: a bounds-checked DMA that clamps the request length to the destination buffer size before \code{mram\_read}. This removes the overflow primitive at its source, so the same exploit leaks the secret 0 of 10 while benign requests complete normally 10 of 10 (\Cref{tab:mitigation}). It is a per-call source fix rather than a platform-wide defense, but it shows the attack is preventable once the DMA length is validated.

\begin{table}[t]
  \centering
\caption{Mitigation experiment ($N{=}10$ per build). Leak = \code{secret} in \code{output\_buf}; caught = canary trip; clean = benign completion.}
  \label{tab:mitigation}
  \begin{tabularx}{\linewidth}{@{}Xccc@{}}
    \toprule
    \textbf{Build} & \textbf{Leak} & \textbf{Caught} & \textbf{Clean} \\
    \midrule
    Baseline (no mitigation)          & 10/10 & ---   & --- \\
    Stack canary (runtime supplied)   & 10/10 & 0/10  & --- \\
    Bounds-checked DMA                 & \textbf{0/10} & --- & 10/10 \\
    \bottomrule
  \end{tabularx}
\end{table}

\subsection{Portability of Transparent Fortification to PIM}
\label{sec:eval:overhead}
The bounds-checked DMA of \Cref{sec:eval:defense} stops the attack on \code{vuln\_dpu} 10 of 10, but \code{vuln\_dpu}'s overflowed buffer is a stack array whose size the compiler can see. Real DPU kernels allocate WRAM through \code{mem\_alloc()}, which returns an opaque pointer with no compile-time size. We asked whether the \code{\_FORTIFY\_SOURCE}-style transparent form of the wrapper---the form a PIM SDK would ship as the default---actually fortifies that code. The question is a defense-portability question rather than a cost question, and we answer it by compiling each kernel twice with the stock upmem toolchain (\code{dpu-upmem-dpurte-clang} 2023.1.0, \code{-O2}, \code{-c}, no linker required): once without the wrapper and once with \code{safe\_mram.h} \code{-include}d so every \code{mram\_read}/\code{mram\_write} macro expands to the fortified form. We then compare the two \code{.text} instruction counts from \code{llvm-objdump~-d} to see whether a bounds check was emitted at all.

\begin{table}[t]
  \centering
    \caption{Whether the transparent \code{safe\_mram.h} bounds-check wrapper actually emits a guard in each kernel.}
  \label{tab:overhead}
  \resizebox{\linewidth}{!}{%
  \begin{tabular}{@{}lrrrrl@{}}
    \toprule
    \textbf{Kernel} & \textbf{Base ins} & \textbf{Hard ins} & \textbf{$\Delta$} & \textbf{Sites} & \textbf{Guard?} \\
    \midrule
    \code{vuln\_dpu}  & 130  & 132  & \textbf{+2} & 1 & yes \\
    \midrule
    VA        & 78    & 78    & 0 & 3 & \textbf{no} \\
    RED       & 114   & 114   & 0 & 1 & \textbf{no} \\
    HST-L     & 159   & 159   & 0 & 3 & \textbf{no} \\
    GEMV      & 190   & 190   & 0 & 7 & \textbf{no} \\
    SpMV      & 148   & 148   & 0 & 5 & \textbf{no} \\
    TRNS      & 394   & 394   & 0 & 7 & \textbf{no} \\
    BS        & 167   & 167   & 0 & 6 & \textbf{no} \\
    TS        & 215   & 215   & 0 & 5 & \textbf{no} \\
    SEL       & 1169  & 1169  & 0 & 2 & \textbf{no} \\
    UNI       & 1382  & 1382  & 0 & 2 & \textbf{no} \\
    SCAN\mbox{-}RSS  & 2152  & 2152  & 0 & 3 & \textbf{no} \\
    NW        & 27    & 27    & 0 & 8 & \textbf{no} \\
    \bottomrule
  \end{tabular}%
  }
\end{table}

\Cref{tab:overhead} reports the measurement, and the headline is a portability finding rather than a cost one: \emph{the transparent host-style bounds-check wrapper is silently inactive on 12 of 12 PrIM kernels we measured.} The reason is specific to how PrIM allocates WRAM. Every DPU kernel in the suite obtains its scratchpad buffers through \code{mem\_alloc()}, which returns an opaque pointer, and \code{\_\_builtin\_object\_size} is unable to attribute a size to such a pointer and returns \code{(unsigned long)-1}. The wrapper itself short-circuits when the size is unknown, so on PIM it skips the check everywhere and the DMA ships unguarded. This is not a toolchain bug; it is exactly how \code{\_FORTIFY\_SOURCE} behaves on glibc when the destination size is unknown. The difference is that on a PIM runtime the ``destination size is unknown'' branch is not an exception but the default, because the standard allocation idiom hides it.

The only kernel in our set where the transparent form fires is \code{vuln\_dpu}, whose overflowed buffer is a 32-byte stack array that \code{\_\_builtin\_object\_size} resolves without help. There the wrapper emits a single \code{jgeu r0, 257, <fail>} (DPU's combined compare-and-branch) plus one extra instruction to materialize the fail-handler target: a two-instruction increase, 1.5\% of the 130-instruction baseline. This is the measured per-site cost of a bounds check when it \emph{is} emitted, and the number useful for projecting the cost of a working defense.

Making bounds-checked DMA a platform-wide default on PIM therefore requires one of two SDK changes, neither of which is in the current runtime. A \code{mem\_alloc}-shaped allocator that returns a size-carrying descriptor rather than an opaque pointer would let the transparent wrapper resolve \code{dst\_size} and emit the check at every call site, which is the natural generalization of glibc's pattern. Alternatively, the kernel-side convention could be changed to declare WRAM buffers as fixed-size arrays so that \code{sizeof()} supplies the size, which is what \code{safe\_mram.h}'s \code{\_bounded} form already exploits. In either case, the per-site cost is the same two instructions we measure on \code{vuln\_dpu}. Scaling from the six runtime-length sites across PrIM identified in \Cref{sec:threat}, the IRAM budget for a working platform-wide default is at most twelve added instructions across the entire suite, negligible against the 4{,}096-instruction IRAM ceiling. Direct wall-clock confirmation on physical upmem silicon remains future work (\Cref{sec:disc:limitations}).

\subsection{Bypassing a Correctly-Placed Canary}
\label{sec:eval:canary-bypass}
A natural response to \Cref{sec:eval:defense} is to assume that a one-time toolchain patch---moving the canary slot to the saved-return side of the frame---would close the gap. We tested this directly. To simulate the corrected placement without rebuilding the compiler, we modified \code{parse\_request} to wrap its vulnerable buffer and a canary word in a single struct, so the canary is guaranteed by the C object model to sit immediately above the buffer and below the saved \code{\{r22,r23\}} pair. The function initializes the canary on entry, re-reads it through a \code{volatile} pointer after the DMA, and calls \code{\_\_stack\_chk\_fail} on mismatch. A linear overflow from the buffer must now cross the canary slot before it can reach the saved return.

We then fired three payloads at the same hardened kernel, built so \code{win()} is physically absent from the binary. The baseline sink-free chain of \Cref{sec:eval:sinkfree} serves as the first case. The second is a canary-aware variant that places the four bytes of the fixed guard value (\code{0x0BADC0DE}, recovered from the DPU runtime binary with a single \code{llvm-objdump} invocation) at the payload offset that lands on the canary slot and continues the saturated \{\reg{r23},\reg{r22}\} pattern above it. The third is the same canary-aware builder with the preservation disabled, as a negative control.

The results are reported in \Cref{tab:canary-bypass}. The baseline chain, which has no knowledge of the guard, trips the canary and halts on every one of 10 runs with no leak. The canary-aware chain leaks the secret on 10 of 10 runs with the canary silent on every one. The DPU reports \code{dpu\_launch status: success}, so the chain reaches \code{\_\_sys\_end} and exits cleanly. The negative control, which clobbers the canary slot with the saturated pattern instead of the guard, trips the canary 10 of 10. The entire bypass costs the attacker four bytes of extra payload at a precomputed offset, and \code{win()} is never reached because it does not exist in this build.

The root cause is not the placement but the \emph{secrecy} of the guard. \code{dpu\_canary.c} fixes \code{\_\_stack\_chk\_guard} at a compile-time constant because the DPU has no entropy source exposed to kernels. A per-launch random guard seeded by the host would close the static-disclosure path we exercise here, but the guard still lives in WRAM at a fixed address with no memory protection, so any attacker primitive that can read or overwrite that word reduces the problem to the static case. We return to this in \Cref{sec:disc:mitigations}.

\begin{table}[t]
  \centering
  \caption{Canary bypass experiment ($N{=}10$ per case). ``leak'' = \code{secret} reached \code{output\_buf}; ``caught'' = \code{\_\_stack\_chk\_fail} marker.}
  \label{tab:canary-bypass}
  \begin{tabularx}{\linewidth}{@{}Xcc@{}}
    \toprule
    \textbf{Payload} & \textbf{Leak} & \textbf{Caught} \\
    \midrule
    Baseline sink-free chain, no canary knowledge  & 0/10  & 10/10 \\
    Canary-aware sink-free chain
      (preserves \code{0x0BADC0DE})                & \textbf{10/10} & 0/10 \\
    Canary-aware builder, preservation disabled
      (negative control)                           & 0/10  & 10/10 \\
    \bottomrule
  \end{tabularx}
\end{table}

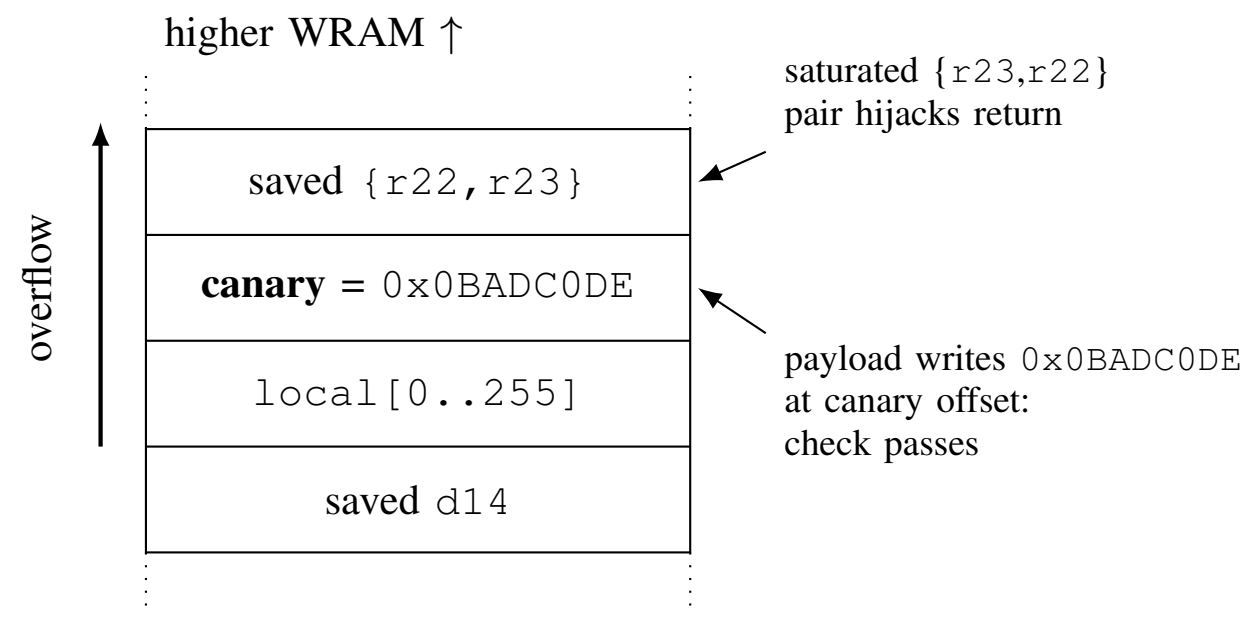
\begin{figure}[t]
\centering
\begin{tikzpicture}[font=\footnotesize, >={Latex[length=1.6mm]}]
  \draw (0,0) rectangle (3.6,2.8);
  \draw (0,0.7) -- (3.6,0.7);
  \draw (0,1.4) -- (3.6,1.4);
  \draw (0,2.1) -- (3.6,2.1);
  \draw[dotted] (0,2.8) -- (0,3.2);  \draw[dotted] (3.6,2.8) -- (3.6,3.2);
  \draw[dotted] (0,0)   -- (0,-0.4); \draw[dotted] (3.6,0)   -- (3.6,-0.4);
  \node at (1.8,0.35) {saved \texttt{d14}};
  \node at (1.8,1.05) {\texttt{local[0..255]}};
  \node at (1.8,1.75) {\textbf{canary} = \texttt{0x0BADC0DE}};
  \node at (1.8,2.45) {saved \texttt{\{r22,r23\}}};
  \node[anchor=west] at (0,3.42) {higher WRAM $\uparrow$};
  \draw[->,thick] (-0.3,0.7) -- (-0.3,2.85);
  \node[rotate=90,anchor=south] at (-0.5,1.75) {overflow};
  \draw[->] (4.1,1.45) -- (3.65,1.75);
  \node[anchor=north west,align=left,font=\scriptsize] at (4.1,1.5)
    {payload writes \texttt{0x0BADC0DE}\\ at canary offset:\\ check passes};
  \draw[->] (4.1,2.65) -- (3.65,2.45);
  \node[anchor=south west,align=left,font=\scriptsize] at (4.1,2.65)
    {saturated \{\texttt{r23},\texttt{r22}\}\\ pair hijacks return};
\end{tikzpicture}

\caption{Bypassing a correctly-placed canary: the attacker preserves the fixed guard \code{0x0BADC0DE} at the canary offset and overwrites the saved return above it. Measured 10/10 in \Cref{tab:canary-bypass}.}
\label{fig:canary-bypass}
\end{figure}

\subsection{Mitigations}
\label{sec:disc:mitigations}
The experiments of \Cref{sec:eval:defense,sec:eval:canary-bypass} together argue that shipping \code{\_\_stack\_chk\_*} in the DPU runtime is necessary but not sufficient: making stack-smashing protection effective on this upward-growing frame requires both a correct canary slot placement (a \llvm{} backend pass over the DPU stack-protector slot ordering) and a host-seeded random guard with a WRAM protection story to prevent recovery. 

The bounds-checked DMA of \Cref{sec:eval:defense} is a small immediately deployable fix (safe \code{mram\_read}/\code{mram\_write} wrappers in the SDK), but \Cref{sec:eval:overhead} shows the \code{\_FORTIFY\_SOURCE}-style transparent form is silently inactive on 12 of 12 PrIM kernels. Making it platform-wide requires either a size-carrying \code{mem\_alloc} descriptor or fixed-size WRAM arrays. Beyond this, a $W\oplus X$ policy for IRAM and a lightweight CFI validating \reg{r23} and \code{jump~rN} targets would raise the bar against code-reuse attacks not tied to a single overflowable buffer.

\paragraph{A hardware-friendly CFI sketch} The structural weakness is that every function return reads its target from attacker-overwritable stack memory. A per-tasklet shadow return-index register would close this gap, in the spirit of Intel's hardware shadow stack in CET~\cite{intel_cet}. It is a small hardware stack of 32-bit instruction indices, separate from the WRAM frame. On a \code{call}, the hardware pushes the next instruction index in parallel with writing \reg{r23}. On \code{jump r23}, it compares the return target against the shadow top and faults on mismatch. Our attacks fail at every chain-hop because the attacker-written \reg{r23} is not the index the caller pushed. The storage cost is modest. Each DPU needs 24 tasklets $\times$ 24 entries $\times$ 4 bytes $=2{,}304$~bytes of SRAM, a first-order estimate under $0.03$~mm$^2$ at a modern $\sim$0.1\,$\mu$m$^2$ 6T-SRAM cell~\cite{chang2013sram}. Full RTL evaluation remains future work, including whether the check fits in fetch slack or requires an added pipeline stage.


The functional simulator does not model cycle timing. We therefore bound the primitive's overhead using static instruction counts from the same hardened PrIM builds in \Cref{tab:overhead}. The relevant dynamic events are \code{call} and \code{jump r23}, which exercise the shadow stack. Across the twelve PrIM kernels, their combined share of static \code{.text} has a median near 8\%.

The hardware primitive adds at most one pipeline cycle per return event. Its dynamic cost is therefore bounded by the return fraction times CPI, which comes out under 8\% on the median kernel. A conservative software emulation of the same semantics would add five extra instructions per round trip, inflating the committed \code{.text} by roughly 40\% at the median. Both numbers are strict overestimates, since hot code paths amortize better than static counts suggest. Both are also strictly worse than a hardware shadow stack, consistent with prior measurements comparing software and hardware shadow-stack costs~\cite{dang2015shadowstacks}.

We report bounds rather than a cycle-accurate measurement because the \upmem{} functional simulator exposes only committed-instruction counts~\cite{upmemsdk}, and a cycle-level DPU model is itself future work (\Cref{sec:future}). Even these loose bounds place the primitive within what a throughput-oriented PIM core can absorb.

The remaining register-indirect terminators, \code{jump r1} and \code{jump r3}, which our scanner catalogs as JOP sinks, require a complementary coarse-grained target check such as a landing-pad instruction at every indirect-call target~\cite{intel_cet}. The combined mechanism is a plausibly affordable in-memory CFI and would turn every chain in this paper into a fault, which is the direction we think a hardened DPU should take.

\subsection{Limitations}
\label{sec:disc:limitations}
Our results come from the \upmem{} functional simulator rather than a physical DIMM. The functional model reproduces the decoded instruction stream and control flow faithfully~\cite{gomezluna2022prim,upmemsdk}, and the toolchain under x86 emulation on an aarch64 host changes tool speed only. Our scanner parses \code{llvm-objdump} text rather than decoding raw bytes, so a gadget the disassembler fails to render would be missed; this makes our counts a lower bound. The published ISA describes 48-bit instructions, yet the ELF encoding is 8-byte aligned, and it is this 8-byte stride our aligned scan relies on.

\subsection{Ethical Considerations}
\label{sec:disc:ethics}
This work was a controlled self-contained study on the \upmem{} functional simulator inside a disposable VM; every program it attacks is intentionally vulnerable code we wrote. Our intent is defensive, and we pair the attack with the mitigations of \Cref{sec:disc:mitigations}. The behaviors we rely on (absent canary runtime, lack of ASLR/CFI) are documented design characteristics of the current platform rather than secret flaws, so no coordinated disclosure timeline is needed.



\section{Future Work}
\label{sec:future}
The natural next offensive step is to automate the manual sink-free chain of \Cref{sec:eval:sinkfree}: a synthesizer that consumes the scanner's gadget JSON and emits a working payload, together with a JOP dispatcher over the \code{jump r1}/\code{jump r3} terminators we catalog. RiscyROP~\cite{cloosters2022riscyrop} does this for RISC-V/ARM64 via symbolic execution; porting that style to the DPU's instruction-index returns and \reg{r22}-threaded reload is the concrete next step. The application interface should also be stressed intra-DPU (24 tasklets share one WRAM address space) and across other PIM designs (Samsung HBM-PIM, SK~Hynix AiM). On the defensive side, the hardware shadow-return-index primitive of \Cref{sec:disc:mitigations} needs RTL prototyping with a cycle-level DPU simulator; all of this ultimately needs validation on real upmem silicon, which would exercise the real MRAM-to-WRAM DMA engine and expose multi-DPU and 24-tasklet concurrency effects the functional model abstracts away.

\section{Conclusion}
\label{sec:conclusion}
We present PIMROP, to our knowledge the first demonstrated code-reuse attack on a commercial Processing-in-Memory ISA, and Galileo-for-DPU, a gadget scanner adapted to the UPMEM DPU's fixed-width, 8-byte-aligned RISC encoding. Using it, we show that Shacham's feasibility thesis holds on real DPU code. The runtime library alone yields 517 non-trivial gadgets spanning all five canonical ROP primitive classes, and every one of 13 PrIM benchmarks carries gadgets as well. We demonstrate two verified end-to-end attacks on the functional simulator. In the first, a stack buffer overflow overwrites the saved return and diverts control into a routine that copies a secret into host-readable memory. In the second, we leak the same secret using a chain of reused gadgets alone, without diverting control into any such routine. We further demonstrate the two reused control-flow primitives needed to lift these chains from fixed straight-line exfiltration to arbitrary computation: a data-dependent conditional branch via a store-into-chain gadget, and a data-dependent loop via ret2loop into \code{strcat}, each measured as deterministic across input variations and together supplying the computational basis for a Turing-complete gadget set on the DPU. We also report two DPU-specific exploitation observations in that the saved return is stored as an instruction index, and the upward-growing stack frame bypasses conventional stack canaries. Moving the canary slot to the correct side of the frame via source-level simulation does not close the gap either: a sink-free gadget chain that preserves the DPU's fixed, host-unseeded guard value leaks the secret on 10 of 10 runs against a kernel with a correctly-placed canary. We also verify a bounds-checked DMA that stops the attack. To enable further research on PIM security, we open-source Galileo-for-DPU and our testing harness at \url{https://github.com/wtegge2/ROP-in-PIM}.

\bibliographystyle{IEEEtran}
\bibliography{references}

\end{document}

%% file: comments.tex
\usepackage{ifthen}
\usepackage{xcolor}
\newboolean{showcomments}
\setboolean{showcomments}{true}

\makeatletter
\newcommand{\mynote}[3]{%
  \ifthenelse{\boolean{showcomments}}{%
   \fbox{\bfseries\sffamily\scriptsize#1}%
   {\small$\blacktriangleright$\textsf{\emph{\color{#3}{#2}}}$\blacktriangleleft$}}%
  {%
   \@bsphack
   \@esphack
  }%
}
\makeatother

\definecolor{asparagus}{rgb}{0.53, 0.66, 0.42}

\newcommand{\fsz}[1]{\mynote{FSZ}{#1}{orange}}
\newcommand{\will}[1]{\mynote{WILL}{#1}{red}}

%% file: authors.tex
\author{%
\IEEEauthorblockN{William C. Tegge}
\IEEEauthorblockA{Syracuse University\\
\texttt{wtegge@syr.edu}}
\and
\IEEEauthorblockN{Shouzhi Fang}
\IEEEauthorblockA{Syracuse University\\
\texttt{sfang18@syr.edu}}
\and
\IEEEauthorblockN{Alex K. Jones}
\IEEEauthorblockA{Syracuse University\\
\texttt{akj@syr.edu}}}
 